# Polarizing oxygen vacancies in insulating metal oxides under high electric field[§]

Mostafa Youssef,[1#] Krystyn J. Van Vliet,[1,2*] and Bilge Yildiz[1,3*]

[1]Department of Materials Science and Engineering

[2]Department of Biological Engineering

[3]Department of Nuclear Science and Engineering, Massachusetts Institute of Technology,

77 Massachusetts Avenue, Cambridge, Massachusetts 02139, USA

#Present Address: Department of Mechanical Engineering, The American University in Cairo, AUC Avenue, P.O. Box 74, New Cairo 11835, Egypt

*Correspondence to: krystyn@mit.edu and byildiz@mit.edu

**ABSTRACT**

We demonstrate a thermodynamic formulation to quantify defect formation energetics in an insulator under high electric field. As a model system, we analyzed neutral oxygen vacancies (color centers) in alkaline-earth-metal binary oxides using density functional theory, Berry phase calculations, and maximally localized Wannier functions. Work of polarization lowers the field-dependent *electric* Gibbs energy of formation of this defect. This is attributed mainly to the ease of polarizing the two electrons trapped in the vacant site, and secondarily to the defect induced reduction in bond stiffness and softening of phonon modes. The formulation and analysis have implications for understanding the behavior of insulating oxides in electronic, magnetic, catalytic, and electrocaloric devices under high electric field.

Growing interest in understanding effects of large electric fields on the polarization, thermodynamics and kinetics of defects in insulating oxides is driven by emerging technologies including resistive switching memories [1,2], electrocaloric refrigeration [3], field assisted ceramic sintering [4], and controlling nanowire growth [5]. Additionally, giant electric fields on the order of 10 MV/cm arise naturally at oxide hetero-interfaces [6,7]. Point defects, particularly oxygen vacancies, play a prominent role in creating interfacial electric fields [8,9] and dictating the functional properties of these metal oxides [10]. The polarization response and thermodynamics of a defect-free insulating crystal under high electric field is well formulated [11–13]. However, the analogous high field effect on a defective crystal remained challenging to address [1,14,15].

Applying a homogeneous electric field $\vec{\mathbf{E}}$ to an insulating crystal bends its electronic bands linearly, and polarizes the crystal uniformly. Thermodynamically, the former effect augments the differential of the internal energy of the crystal $dU$ by a charge transfer or electrochemical work $\phi dq$ [12]. Here, $\phi$ is the electrostatic potential and $q$ is the charge transferred. The second effect extends $dU$ by what is known as the polarization work $\vec{\mathbf{E}} \bullet d(V\vec{\mathbf{P}})$, where $V$ is the crystal volume and $\vec{\mathbf{P}}$ is its macroscopic polarization [12]. A perfect crystal is not affected by $\phi dq$ since it is neutral. On the contrary, charged defect equilibria in an insulating defective crystal are affected strongly by $\phi dq$. This electrochemical effect has been exploited to control the defect equilibria in $CeO_2$ [16] and phase transitions in $SrCoO_x$ [17]. In contrast, polarization work is well analyzed for perfect crystals [18,19] and was invoked to predict electric field effect on the phase diagram of defect-free water [20] (ions are the defects of liquid water [21]) and on the phase transitions of defect-free $HfO_2$ and $ZrO_2$ [22]. However, there is no detailed and quantitative analysis for the impact of polarization work on a realistic insulator that contains point defects. In particular, we seek a thorough analysis that spans from the global effects of electric field on the abundance of defects, down to the local effects on the single defect site. In this letter, we adopt the neutral oxygen vacancy $V_O^\times$ in MgO, CaO, SrO, and BaO as a model system to study polarization effects. This class of oxides is important due to their abundance on Earth [23], and their potential use in catalysis [24], electronics [25] and even as ferroelectrics [26]. The study of this neutral defect allows us to focus on polarization effects, as we intentionally preclude any contribution from electrochemical work. This defect, which is also known as the color center, is the canonical intrinsic defect in these oxides [27].

In this Letter, using density functional theory (DFT) and modern theory of polarization [28] we reveal that the abundance of $V_O^\times$ is enhanced by the work of polarization. We attribute this enhancement to two factors; primarily the ease of polarizing the two electrons trapped in $V_O^\times$, and secondarily the softening of some phonon modes and reduction in stiffness of bonds in the defective crystal containing $V_O^\times$. These conclusions are supported by analyzing the polarization field of the defect, and the static dielectric permittivities of both the perfect and defective crystals.

For an insulating metal oxide under electric field, the first differential of internal energy is:

$$dU = TdS - PdV + \mu_O dN_O + \sum_k \mu_k dN_k + \mu_e dn_e + \phi dq + \vec{\mathbf{E}} \bullet d(V\vec{\mathbf{P}}), \qquad (1)$$

where $T$, $S$ and $P$ are the temperature, entropy, and pressure, respectively. The chemical potentials $\mu_O$, $\mu_k$, and $\mu_e$ are those of oxygen, cation $k$, and electrons, respectively; and $N_O$, $N_k$, $n_e$, are the number of particles of oxygen, cation $k$, and electrons, respectively. The summation is taken over all types of cations in the oxide. A partial Legendre transform of $U$ provides a convenient expression in terms of natural variables that can be varied experimentally such as *T*, *P*, $\mu_O, \phi$, and $\vec{\mathbf{E}}$ [29]. Moreover, for theoretical convenience in treating charged defects, the transform also includes $\mu_e$ as a natural variable. We define the resulting thermodynamic potential as the *electric* Gibbs free energy and denote this by $G_E$:

$$G_E = U - TS + PV - \mu_O N_O - \mu_e n_e - \phi q - V\vec{\mathbf{E}} \bullet \vec{\mathbf{P}}. \quad (2)$$

Here, we restrict the analysis to $T = 0$ K, assume no electrostriction (hence $\Delta V = 0$), and consider neutral defects (hence $\Delta q = 0$). In addition, following the arguments in reference [20] we do not consider depolarization fields, and as such $\vec{\mathbf{E}}$ is the applied external field. Under such assumptions we define the electric Gibbs energy of formation, $G_E^{form}$, of the neural defect $V_O^{\times}$ to be:

$$G_E^{form} = (U^{def} - U^{perf} + \mu_O) - V\vec{\mathbf{E}} \bullet (\vec{\mathbf{P}}^{def} - \vec{\mathbf{P}}^{perf}), \quad (3)$$

where the superscripts *def* and *perf* denote the defective and perfect crystals, respectively. The first term is the defect formation energy, $U^{form}$. The second term in Eq. (3) is the polarization work of primary interest herein, where we identify $V(\vec{\mathbf{P}}^{def} - \vec{\mathbf{P}}^{perf})$ as the defect dipole moment, $\vec{\mathbf{p}}_{V_O^{\times}}$. In fact, $U^{form}$ under constant electric displacement field ($\vec{\mathbf{D}}$), which corresponds to open-circuit boundary conditions [19], has been computed previously for neutral defects in thin film Si [30] and $TiO_2$ [31] using a sawtooth potential. However, under constant $\vec{\mathbf{E}}$ which corresponds to closed-circuit boundary conditions [19], $G_E^{form}$ is the relevant thermodynamic potential, and thus the work of polarization is crucial for accurate description of defect thermodynamics under high $\vec{\mathbf{E}}$. (See Supplemental Material (SM) [32] section 1.d for more details.)

We calculated the responses of rock-salt MgO, CaO, SrO, and BaO to external electric fields using DFT and Berry phase approach [33,34] as implemented in the QUANTUM ESPRESSO package [35]. Ultrasoft pseudopotentials [36–38] represented the interaction between core and valence electrons and the revised Perdew, Burke, and Ernzerhof functional for solids (PBEsol) [39] described the exchange correlation. $\vec{\mathbf{E}}$ was applied along the cation-oxygen bonds in [100] direction. By removing the arbitrariness in the polarization quantum, we identified the correct polarization branch for each of the perfect and defective crystals, and thereby quantified the work of polarization in Eq. (3) for formation of $V_O^{\times}$. To analyze the local polarization field surrounding the defect site, we invoke the well-established relationship between Wannier centers and polarization [28,40]. Thus, we computed maximally localized Wannier functions [40] from the original polarized Bloch states using the software WANNIER90 [41]. Further details are included in SM [32].

The field dependence of the relative $G_E^{form}$ of $V_O^{\times}$ in the four oxides is shown in FIG. 1(a). $\Delta G_E^{form}$ decreases monotonically in all cases, though more pronounced in BaO. In FIG. 1(b) the dependence of $\Delta U^{form}$ is shown, and indicates a monotonic increase in MgO, CaO, and SrO, but an initial increase followed by a decrease for $|\vec{\mathbf{E}}|$ > 3 MV/cm in BaO. This behavior of $\Delta U^{form}$ is attributable to the static permittivities of the defective and perfect crystals as discussed later. The fact that $\Delta G_E^{form}$ does not follow the behavior of $\Delta U^{form}$ shows clearly the importance of the polarization work term in Eq. (3), which favors the formation of the defect with increasing electric field by lowering $\Delta G_E^{form}$.

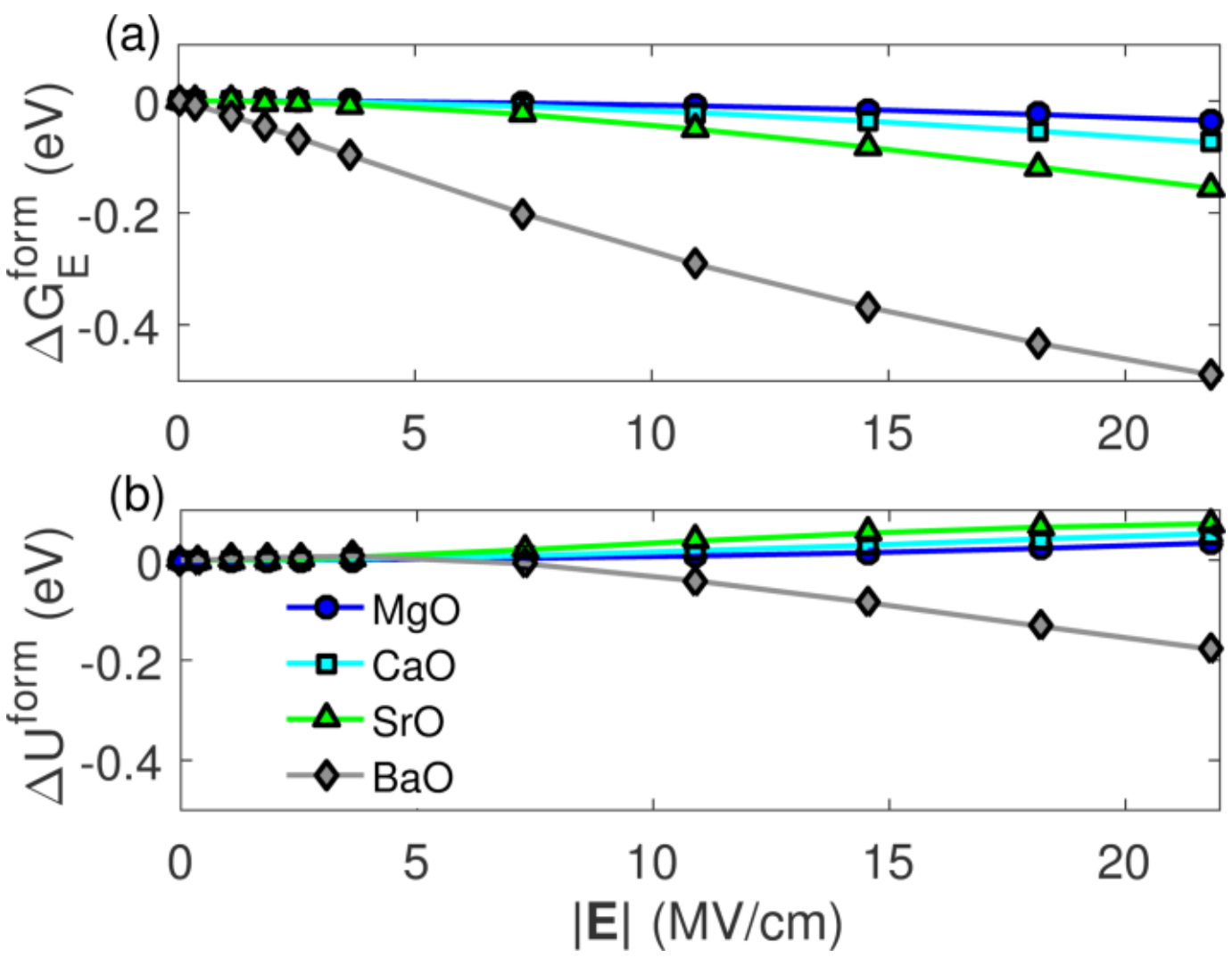

FIG. 1. (a) Relative electric Gibbs free energy of formation and (b) relative formation energy of $V_O^{\times}$ as a function of electric field in the studied oxides.

As simple dielectrics, the four oxides exhibit linear $|\vec{\mathbf{P}}|-|\vec{\mathbf{E}}|$ relationships (Fig. S1 in SM [32]). Nonlinearities arise due to defects. FIG. 2(a) shows the field-dependent dipole moment of $V_O^{\times}$ in units of Debye (D). To provide a convenient reference for polarity, we also show the zero-field gas-phase dipole moment of the highly polar water molecule, $|\vec{\mathbf{p}}_{H_2O}^{0}|$ of magnitude 1.86 D [42]. At zero-field, $|\vec{\mathbf{p}}_{V_O^{\times}}^{0}|$=0 as dictated by the symmetry of the rock-salt lattice (see SM section 2 [32]). At finite field, both $\vec{\mathbf{P}}^{perf}$ and $\vec{\mathbf{P}}^{def}$ are parallel to $\vec{\mathbf{E}}$. Thus a positive value of $|\vec{\mathbf{p}}_{V_O^{\times}}|$ implies that $|\vec{\mathbf{P}}^{def}|>|\vec{\mathbf{P}}^{perf}|$ and this is the case for the four oxides. In MgO, $|\vec{\mathbf{p}}_{V_O^{\times}}|$ remains linear with $|\vec{\mathbf{E}}|$, and up to the highest field considered here its magnitude remains less than $|\vec{\mathbf{p}}_{H_2O}^{0}|$. Nonlinearity appears in CaO and SrO, in which $V_O^{\times}$ can be as polar as gas-phase $H_2O$ at fields > 11.5 MV/cm and > 4.2 MV/cm, respectively. A more dramatic nonlinearity occurs in BaO where initially $|\vec{\mathbf{p}}_{V_O^{\times}}|$ rises to 7.5$|\vec{\mathbf{p}}_{H_2O}^{0}|$ at a field of 3.6 MV/cm and then reduces but remains positive up to the highest field considered. The initial sharp increase is due to a reduction in the stiffness of some bonds [43] caused by the creation of the defect. The reduction of $\left|\vec{\mathbf{p}}_{V_O^\times}\right|$ at even higher $\left|\vec{\mathbf{E}}\right|$ occurs when the bond stiffness around the defect increases relative to that of the perfect crystal under the electric field. We elaborate more on these aspects below.

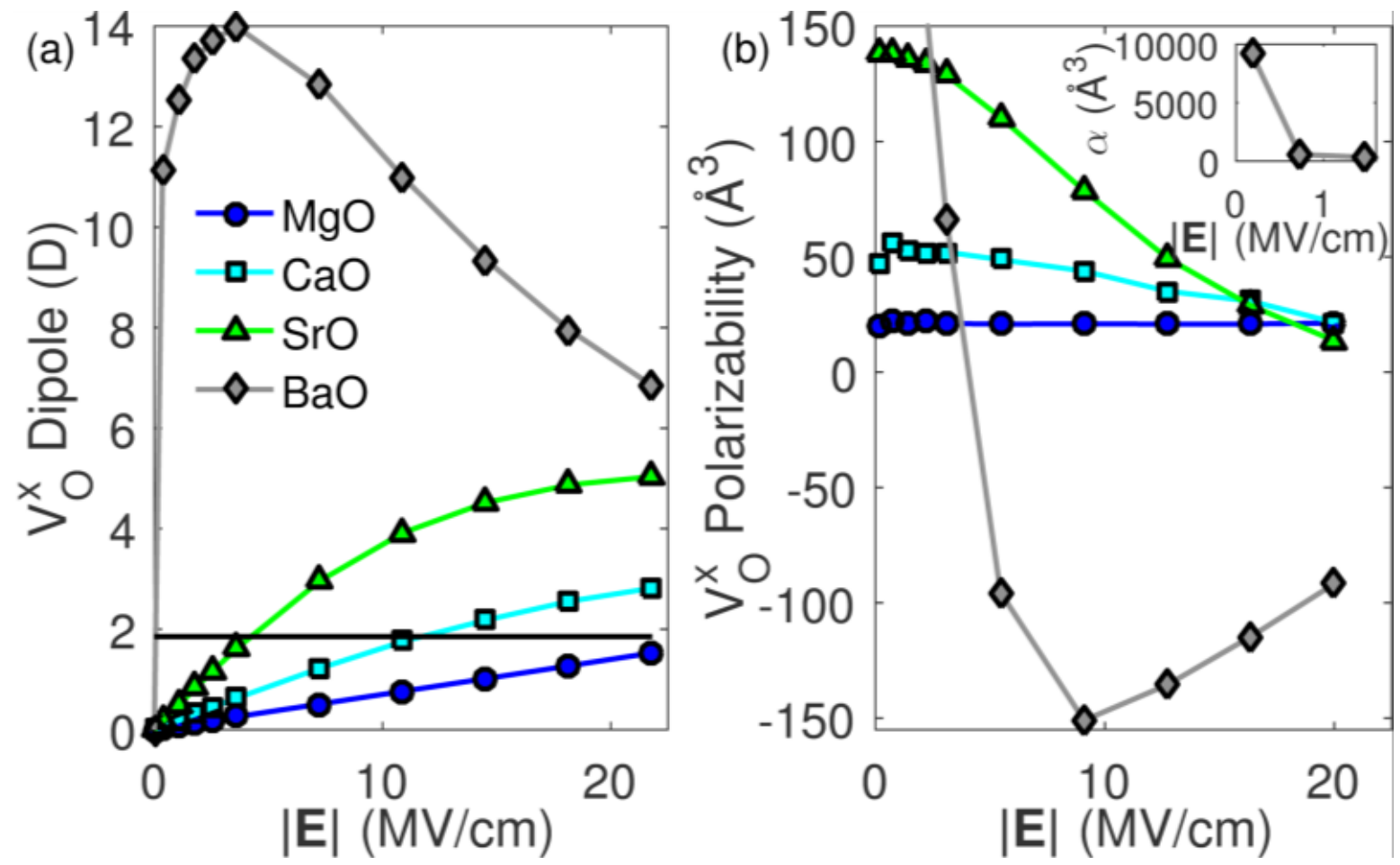

FIG. 2. (a) The field dependence of the dipole moment of $V_O^\times$, $\left|\vec{\mathbf{p}}_{V_O^\times}\right|$. For comparison the zero-field dipole moment of the gas-phase water molecule, $\left|\vec{\mathbf{p}}^0_{H_2O}\right|$=1.86 D [42], is indicated by the black horizontal line. (b) Field dependent polarizability of $V_O^\times, \alpha_{V_O^\times}$ . The inset focuses on the low-field polarizability in the case of BaO.

To describe the spatial distribution of the polarization field around the defect site, we define the polarizability tensor of the defect $\boldsymbol{\alpha} = \partial\vec{\mathbf{p}}_{def} / \partial\vec{\mathbf{E}}$ which is scalar in this work. We note that our definition does not include dipole-dipole interactions [44,45] since we are concerned here with non-interacting defects. The field-dependent polarizability of $V_O^\times$ is presented in FIG. 2(b). Magnitudes of $\alpha$ for $V_O^\times$ under low (zero)-field are 20, 46, 139, and 9175 Å$^3$ in MgO, CaO, SrO, and BaO, respectively, increasing with the size of the host lattice (Section 1.e. in SM [32]). There have been attempts to compute the low-field polarizability for the color center in alkali metal halides using model Hamiltonians, with reported values ranging between 10 and 55 Å$^3$ [46].

The invariance of $\alpha$ for $V_O^{\times}$ in MgO as a function of $\vec{\mathbf{E}}$ mainly reflects the fact that the field stiffens the bonds in both the perfect and defective crystals at the same pace. In contrast, in CaO, SrO, and BaO, $\alpha$ is a decreasing function of $\vec{\mathbf{E}}$, indicating that $\vec{\mathbf{E}}$ stiffens the bonds at a faster pace in the defective crystal. In BaO, $\alpha$ becomes negative when most of the bonds in the defective crystal become stiffer than their counterpart in the perfect crystal as we explain later with FIG. 4.

A natural question emerges from this discussion: why does work of polarization lower $G_E^{form}$ of $V_O^{\times}$? Equivalently, why is the defective crystal more polarized compared to the perfect crystal? We propose two answers. First, $V_O^{\times}$ is essentially a vacant site on the oxygen sublattice, containing two trapped electrons. The absence of the confining potential of the nucleus of the removed oxygen atom, together with the vacant space available to the two trapped electrons, facilitates more extensive polarization of these two electrons compared to the polarization of the oxide ion at this position in the perfect crystal. A similar argument is invoked to explain the larger polarizabilities of ions in the gas-phase relative to those in condensed matter [44,47]. Second, the creation of the vacancy softens some phonon modes and reduces the stiffness of the bonds around the vacancy site. These bonds with reduced stiffness are then more polarizable under electric field. We further support these two arguments with the subsequent analysis.

The two electrons trapped in $V_O^{\times}$ occupy an in-gap state derived from *s*-like orbitals of the surrounding cations (Section 1.f in SM [32]). The zero-field charge densities of these two electrons in the four oxides considered are depicted schematically in FIG. 3(a-d). An electric field applied along the [100] or +*x* direction deforms the charge density of the two electrons such that it is depleted in +*x* and accumulated in –*x* as shown in FIG. 3(e-h) under a field of 21.8 MV/cm. This electronic deformation is minimal in the case of MgO, and is very pronounced in BaO.

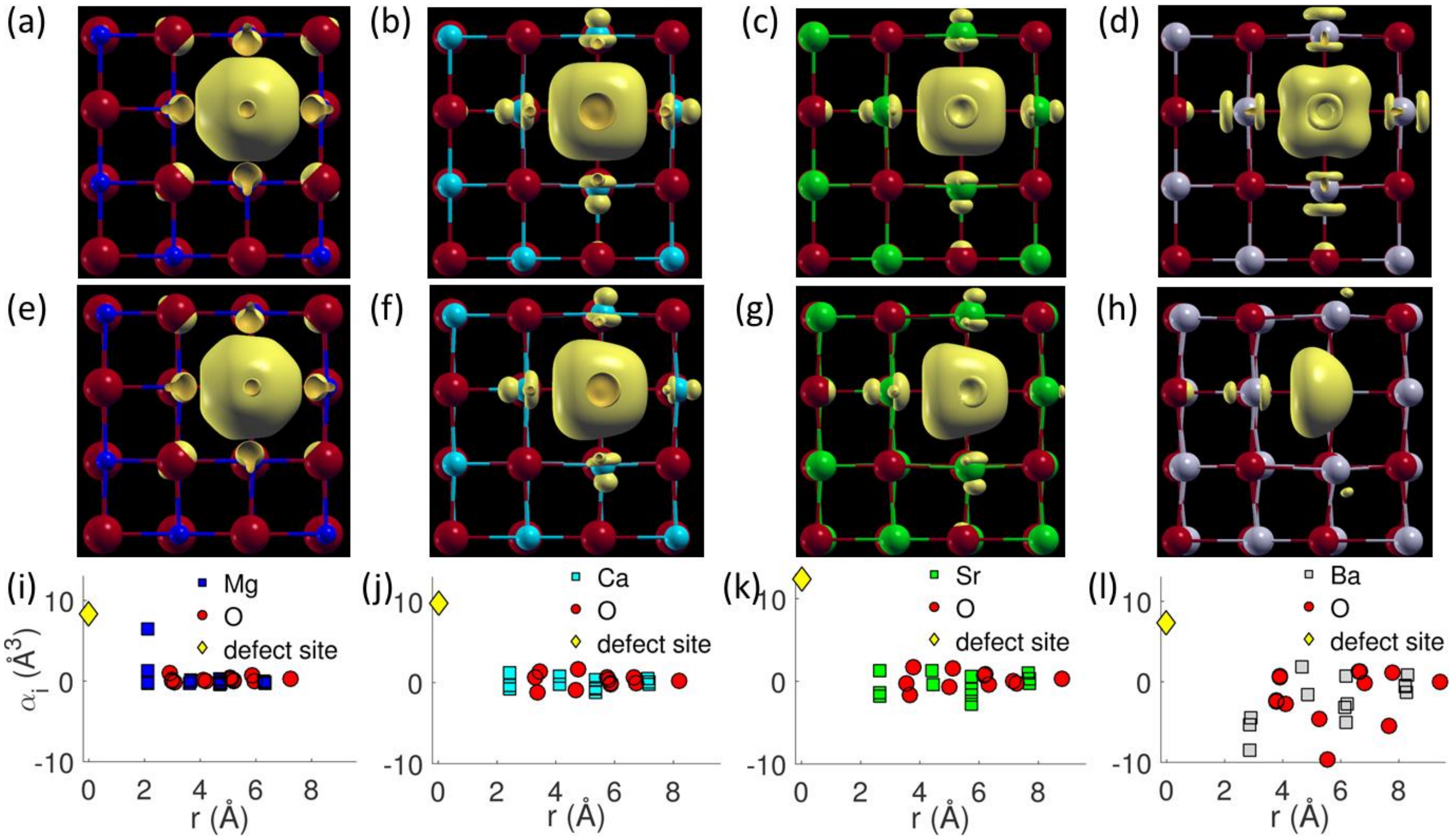

FIG. 3. Visualizations of the charge density of the two electrons trapped in $V_O^{\times}$ at zero field in (a) MgO, (b) CaO, (c) SrO, and (d) BaO. Similar visualizations at a field of 21.8 MV/cm in +*x* direction are shown for (e) MgO, (f), CaO, (g) SrO, and (h) BaO. Red, blue, cyan, green, and grey spheres represent O, Mg, Ca, Sr, and Ba ions, respectively. The yellow isosurfaces in (a-h) represent the electronic charge density and are taken at 15% of the maximum value in each plot. These visualizations were generated using the software XCRYSDEN [48]. (i-l) show high-field site-decomposed polarizability, $\alpha_i$, as a function of distance, *r*, from the defect site, in the case of (i) MgO, (j) CaO, (k) SrO, and (l) BaO. $\alpha_i$'s were calculated by finite difference between field values of 18.2 and 21.8 MV/cm.

To quantify the contribution of each lattice site to the overall defect polarizability, we compute a site-decomposed polarizability $\alpha_i$ by invoking the Wannier centers belonging to this lattice site *i* such that $\alpha_{V_O^\times} = \sum_{i \in supercell} \alpha_i$ (SM section 1c [32]). In FIG. 3(i-l) we present the high-field $\alpha_i$ for the different lattice sites surrounding the defect. Note that $\alpha_i$ at the defect site is the difference between the contribution of the two trapped electrons at the defect site in the defective crystal and the contribution of the oxide ion that occupies the very same site in the perfect crystal. It is evident that major contributors to the polarizability of $V_O^\times$ are the two electrons trapped in the defect site whose high-field $\alpha_i$ are on the order of 10 Å$^3$. Even in BaO when the overall high-field $\alpha$ for $V_O^\times$ is negative, $\alpha_i$ remains positive for the two trapped electrons. This supports our first argument that these two trapped electrons are easier to polarize under electric field in comparison to the oxide ion.

The calculated static permittivities of the perfect crystals $\varepsilon^{perf}$ and defective crystals $\varepsilon^{def}$ are shown in FIG. 4. The low(zero)-field $\varepsilon^{perf}$ for the considered oxides are in reasonable agreement with experimental values [49], with the exception of BaO [49,50] (SM section 3 [32]). The figure also shows that the application of $\vec{\mathbf{E}}$ reduces $\varepsilon$ monotonically for all cases. We attribute this decrease to the reduction in the contribution to $\varepsilon$ from the ionic relaxation because the clamped-ion contribution to $\varepsilon$ is field-independent (SM section 3 [32]). The ionic relaxation contribution is inversely proportional to $\omega_i^{\,2}$, where $\omega_i$ is the angular frequency of the zone-center phonon mode *i* [51]. The field hardens the phonon modes (increases $\omega_i$), and so $\varepsilon$ decreases. FIG. 4 also shows that $\varepsilon^{def}$ is generally greater than $\varepsilon^{perf}$ for all fields with the exception of BaO when $|\vec{\mathbf{E}}|$> 3 MV/cm. $\varepsilon^{def}$ being greater than $\varepsilon^{perf}$ reveals that $V_O^\times$ softens some of the phonon modes and reduces the stiffness of bonds in the defective crystal. Since BaO has the largest lattice constant among the studied oxides, introducing $V_O^{\times}$ brings BaO to the verge of being ferroelectric as evidenced from the large $\varepsilon^{def}$ at low field shown in FIG. 4(b).

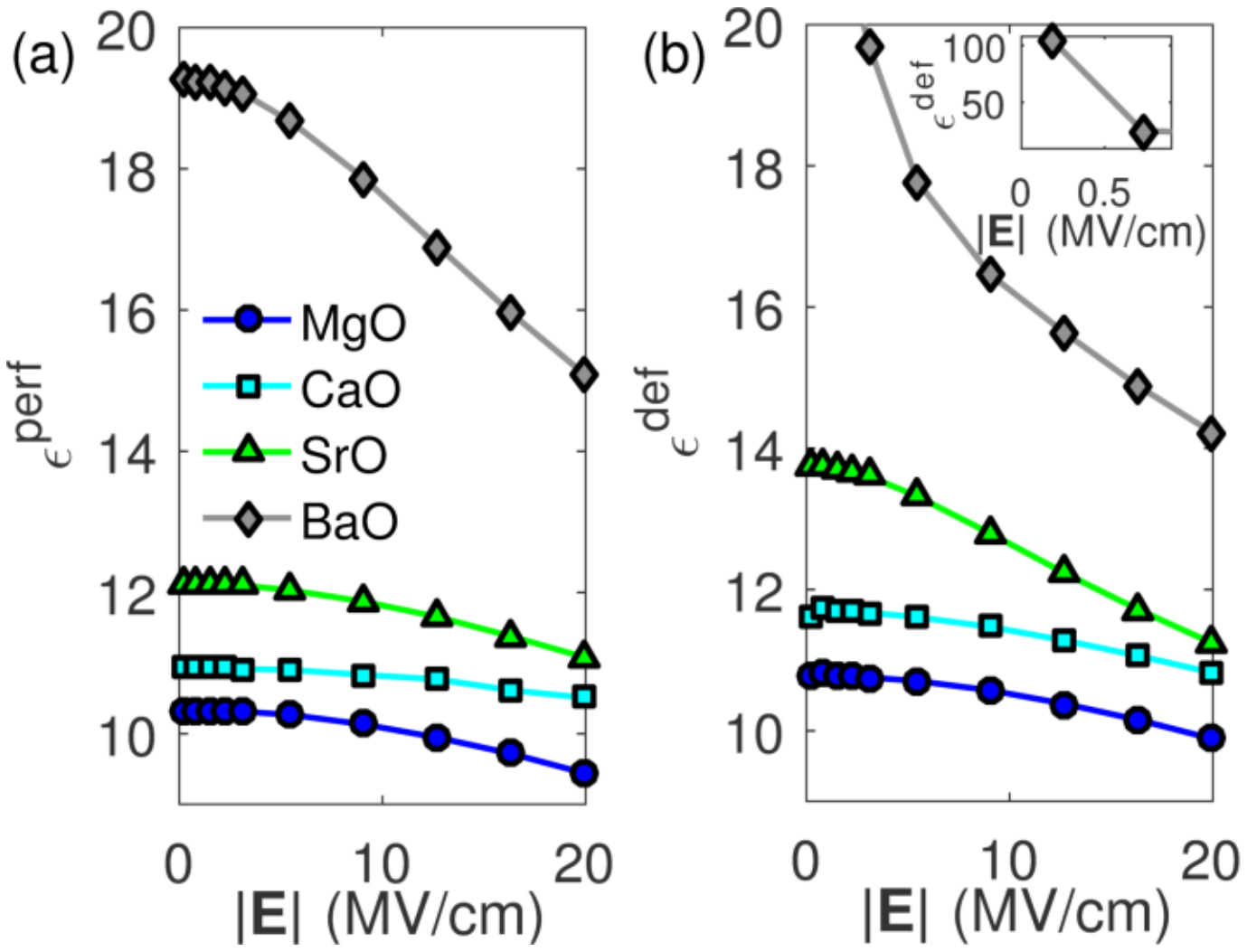

FIG. 4. Field dependent static permittivity of (a) the perfect crystal and (b) the defective crystal containing $V_O^{\times}$ for the studied oxides. The inset in (b) focuses on the $\varepsilon^{def}$ of BaO at low fields.

The reduction in bond stiffness introduced by $V_O^{\times}$ facilitates bond deformation and stores more associated potential energy under $\vec{\mathbf{E}}$. Macroscopically, note from Eq. 1 that $\partial U / \partial \vec{\mathbf{E}} = V(\varepsilon\varepsilon_0 - \varepsilon_0)\vec{\mathbf{E}}$, where $\varepsilon_0$ is the vacuum dielectric permittivity. When $\varepsilon^{def} > \varepsilon^{perf}$, $\Delta U^{form}$ monotonically increases with $\vec{\mathbf{E}}$; this is the case for all of these oxides except BaO at $|\vec{\mathbf{E}}| > 3$ MV/cm, beyond which $\varepsilon^{def}$ becomes less than $\varepsilon^{perf}$. Note that $\varepsilon^{def} - \varepsilon^{perf}$ is essentially the defect polarizability (FIG. 2(b)) scaled by the crystal volume, $\alpha / V$. Microscopically and using a harmonic approximation, the energy stored in a bond is ½ $k\Delta x^2$, where $k$ is the bond stiffness and $\Delta x$ is the bond deformation. Since $V_O^{\times}$ reduces the stiffness of some of the bonds, this increases $\Delta x$ of these bonds under $\vec{\mathbf{E}}$ and the overall stored potential energy. This explains both macroscopically and microscopically the behavior of $\Delta U^{form}$ in FIG. 1(b). Since the bonds with reduced stiffness in the defective oxides deform more readily under $\vec{\mathbf{E}}$, this also means that these bonds are more readily polarized under $\vec{\mathbf{E}}$. This supports our second argument related to the defective crystal being more polarized than the perfect crystal which eventually contributes to lowering $G_E^{form}$ of $V_O^{\times}$.

The field itself hardens the phonon modes and increases the bond stiffness in both the perfect and defective crystals. $V_O^{\times}$ on the other hand softens the phonon modes and reduces bond stiffness in the defective crystal, and this effect of $V_O^{\times}$ prevails against the field effect up to the highest field considered here, except for BaO when $|\vec{\mathbf{E}}| > 3$ MV/cm. Since the defective BaO starts with much softer modes compared to the other oxides, the rate of mode hardening under the field is faster [52] for defective BaO and thus at 3 MV/cm both the perfect and defective BaO have *effectively* similar phonon mode frequencies and bond stiffness (See also SM [32] section 5).

Lastly, we emphasize that $\Delta G_E^{form}$ is dictated by the relative polarizability of the defective crystal with respect to that of the perfect crystal. This relative polarizability cannot be expressed simply in terms of Born effective charge $Z^*$ of the cation in the perfect crystal. Although the qualitative order of $\Delta G_E^{form}$ in FIG. 1(a) matches the order $Z_{Mg}^* = +2.0 < Z_{Ca}^* = +2.3 < Z_{Sr}^* = +2.4 < Z_{Ba}^* = +2.7$ that we calculated using density functional perturbation theory [53] for the perfect crystals, this does not necessarily hold for all oxides. We support this understanding by calculating the field-dependent $G_E^{form}$ for $V_O^{\times}$ in cubic $SrTiO_3$ (See

SM section 4 [32] for details and discussion of potential phase transition in $SrTiO_3$) . In spite of the very large $Z_{Ti}^{*} = +6.4$ compared to Ti formal charge of +4 in $SrTiO_3$ and compared to the cations in the binary oxides, the applied field does not lower $G_{E}^{form}$ for $V_{O}^{\times}$ in $SrTiO_3$ to the same extent as it does in BaO. Perfect crystal $SrTiO_3$ is highly polarizable as implied by $Z_{Ti}^{*}$, but so is $SrTiO_3$ containing oxygen vacancies, and the net difference is less than the net difference in polarizability obtained in BaO.

In summary, we investigated the effect of high electric fields on the polarization of neutral oxygen vacancies in alkaline-earth-metal binary oxides. We showed that, beyond the electrochemical effect that is classically null for a neutral defect, the polarization work lowers the *electric* Gibbs energy of defect formation. This was explained by the greater polarizability of the defective crystal compared to the perfect crystal, primarily due to the ease of polarizing the two electrons trapped in the vacant site and due to the reduction in bond stiffness. Accounting for polarization work is necessary for a better understanding of redox based memristive devices. Additionally, our analysis of field-dependent defect polarizability suggests that the assumption of fixed dipoles used in studying electrocaloric refrigerators [54,55] can be relaxed. Future studies can also include implications of defect polarization under electric field on defect diffusion [56].

This work was supported by the MRSEC Program of the National Science Foundation (NSF) under award number DMR – 1419807. This research used resources of the National Energy Research Scientific Computing Center, a DOE Office of Science User Facility supported by the Office of Science of the U.S. Department of Energy under Contract No. DE-AC02-05CH11231. M.Y. thanks Prof. Paolo Giannozzi of University of Udine for helpful comments on Berry phase implementation in QUANTUM ESPRESSO.

# Supplemental Material for

# Polarizing oxygen vacancies in insulating metal oxides under high electric field

Mostafa Youssef,[1] Krystyn J. Van Vliet,[1,2*] and Bilge Yildiz[1,3*]

[1]Department of Materials Science and Engineering

[2]Department of Biological Engineering

[3]Department of Nuclear Science and Engineering, Massachusetts Institute of Technology,

77 Massachusetts Avenue, Cambridge, Massachusetts 02139, USA

*Correspondence to: krystyn@mit.edu and byildiz@mit.edu

List of contents:

## 1. Supplemental methods and theoretical approach.

### 1.a. Details of density functional theory calculations.

All density functional theory (DFT) calculations were performed using the code PWSCF of the QUANTUM ESPRESSO package version 5.2.0 [1]. The plane-wave kinetic energy cutoff was set to 45 Ry and charge density cutoff to 360 Ry. Ultrasoft pseudopotentials [2] generated using the revised Perdew, Burke, and Ernzerhof functional for solids (PBEsol) [3] were selected from two recent pseudopotential libraries; GBRV library of Garrity, Bennett, Rabe, and Vanderbilt [4] and PSlibrary of Dal Corso [5]. Our selection benefited from the extensive verification effort of I.E. Castelli et al. [6]. A list of the pseudopotentials and the number of electrons treated as valence electrons for each of them are summarized in Table S1.

Table S1. List of the pseudopotentials used in this work.

| Element | Valence electrons | Name of the pseudopotential | Library |
|---|---|---|---|
| Mg | 10 | mg_pbesol_v1.4.uspp.F.UPF | GBRV |
| Ca | 10 | ca_pbesol_v1.uspp.F.UPF | GBRV |
| Sr | 10 | Sr.pbesol-spn-rrkjus_psl.1.0.0.UPF | PSlibrary |
| Ba | 10 | ba_pbesol_v1.uspp.F.UPF | GBRV |
| O | 6 | o_pbesol_v1.2.uspp.F.UPF | GBRV |

To evaluate the equilibrium lattice constant of the metal oxides considered in this work (MgO, CaO, SrO, and BaO) according to PBEsol functional and the selected pseudopotentials, we performed a series of constant volume relaxations for each oxide using 29 different volumes. All of the 29 calculations were fit to a 3$^{rd}$ order Birch-Murnaghan [7] equation of state using the code PHONOPY version 1.9.3 [8]. From the fit, the equilibrium lattice constant and bulk modulus of the rock-salt structure of these oxides were extracted. These calculations were performed on a rock-salt conventional cell of each of these oxides (4 formula units in the cell). The reciprocal space was sampled using $4\times4\times4$ displaced Monkhorst-Pack *k*-point grid [9]. No smearing was used in these calculations (fixed occupations). A summary of the equilibrium lattice constant, bulk modulus, and electronic band gap for each oxide accompanied by selected experimental values for these properties are presented in Table S2.

Table S2. Comparison between our DFT-PBEsol calculated values for the lattice constant, bulk modulus, and band gap of the oxides considered in this work and selected experimental values.

| Oxide | Lattice constant (Å) | | Bulk modulus (GPa) | | Band Gap (eV) | |
|---|---|---|---|---|---|---|
| | This work | Experiment | This work | Experiment | This work | Experiment |
| MgO | 4.217 | 4.212 [10] | 157 | 160 [10] | 5.7 | 7.8 [11] |
| CaO | 4.769 | 4.811 [12] | 116 | 112 [12] | 4.3 | 7.1 [11] |
| SrO | 5.126 | 5.160 [13] | 94 | 91 [13] | 3.9 | 5.9 [14] |
| BaO | 5.513 | 5.539 [15] | 78 | 69-72 [15] | 2.6 | 4.8 [16] |

**1.b. Electric field calculations.**

Electric field calculations of the perfect and defective crystals were performed in supercells made of $2\times2\times2$ conventional cells (32 unit formula). Due to the large expense of ionic and electronic relaxations, the reciprocal space in these calculations was sampled using a single *k*-point at (0.5, 0.5, 0.5) of the reciprocal of the supercell. Such off-Gamma sampling is desirable for defect calculations [17]. For each structure (perfect or defective crystal), we performed full relaxations under 11 electric fields at field values of {0, 0.0001, 0.0003, 0.0005, 0.0007, 0.001, 0.002, 0.003, 0.004, 0.005, 0.006} in Ry atomic units. The electric field was applied using Berry phase approach and modern theory of polarization [18,19]. At each field value both electronic and ionic degrees of freedom were allowed to fully relax, however the cell volume and shape were fixed. In other words, electrostriction effects were not considered here. The stopping criterion for electronic relaxation was set to $10^{-8}$ Ry, whereas ionic relaxation stopping criteria were set to $4\times10^{-6}$ Ry for total energy and $4\times10^{-5}$ Ry/bohr for forces. These somewhat strict stopping criteria were found necessary to obtain smooth variations of the net polarization vs. electric field which consequently implies smooth variations of the dielectric constant vs. electric field. All calculations were spin-polarized; however, none of the resulting structures exhibited net magnetic moment. To accelerate the convergence of the electronic structure, especially in the case of defective crystals, we applied a small Gaussian smearing of 0.004 Ry. This appears to be in conflict with using Berry phase approach which strictly requires *integer* electronic occupations, whereas smearing typically leads to *fractional* occupations and metallic-like solutions. In our case, the small smearing we applied, preserved integer electronic occupations and simultaneously achieved noticeably faster convergence of the electronic structure. We performed test cases using fixed occupations and zero smearing and confirmed that the small 0.004 Ry smearing did not introduce any artifacts in the resulting polarization. There is, however, one exception which is the oxygen vacancy in BaO at zero electric field. This particular case deserves more discussion and is presented in section 2 below.

In analyzing the polarization computed using modern theory of polarization there is an uncertainty; polarization is defined modulo a quantum [20]. This typically poses a problem in determining the spontaneous polarization of a ferroelectric material. However, this issue was not of concern here since the oxides considered in this work are all simple dielectrics. Even the neutral oxygen vacancy in these oxides exhibits a symmetric electronic structure and hence does not possess a net dipole moment at zero electric field. In the paper, we presented the relative change in the formation energy and electric Gibbs energy of $V_O^\times$ as a function of electric field. For the record, we present here the absolute formation energy at zero field and under *oxygen rich* conditions in Table S3.

Table S3. The formation energies of neutral oxygen vacancies in the oxides considered in this work in eV. These were calculated under oxygen rich conditions using the DFT energy of the $O_2$ molecule as reference. No corrections applied to account for the overbinding of the $O_2$ molecule typical for DFT functionals [21].

| MgO | CaO | SrO | BaO |
|---|---|---|---|
| 6.79 | 7.37 | 7.21 | 7.00 |

### 1.c. Wannier functions.

Maximally localized Wannier functions (MLWF) [22,23] were calculated using the code WANNIER90 version 1.2 [24] starting from the field-polarized Bloch states generated by the code PWSCF. The initial guess for the projections functions was set to randomly-centered *s*-type Gaussians. As in Berry phase calculations the reciprocal space was sampled using a single *k*-point at (0.5, 0.5, 0.5) of the reciprocal of the supercell. In the wannierization we targeted a number of MLWFs to represent the valence band only in the perfect crystal and both the valence band and the defect state in the case of the defective crystal. Thus no wannierization was performed for the conduction band states. This automatically triggers the disentanglement algorithm of Souza-Marzari-Vanderbilt [25] in WANNIER90. However, the defect state is actually well-separated from the conduction band and no disentanglement is needed (see section 2 below for the special case of BaO at zero field). In all cases we performed 10000 iterations to achieve convergence in the spread of Wannier functions within $10^{-6}$ Å$^2$ but the Wannier centers typically converge with much fewer iterations [22]. The centers are all what we need to calculate the polarization. It is worth noting that the Wannier functions associated with the defect state have shapes that exactly look like the corresponding Bloch states shown in Figure 3 in the paper. The reason is that these defect states are well-isolated from both the conduction and valence bands.

To check the consistency between the total polarizations calculated using Berry phase approach (PWSCF) and Wannier centers approach (WANNIER90) we plot the polarization-field relationship in the perfect crystals and defective crystals using both methods in Figure S1.

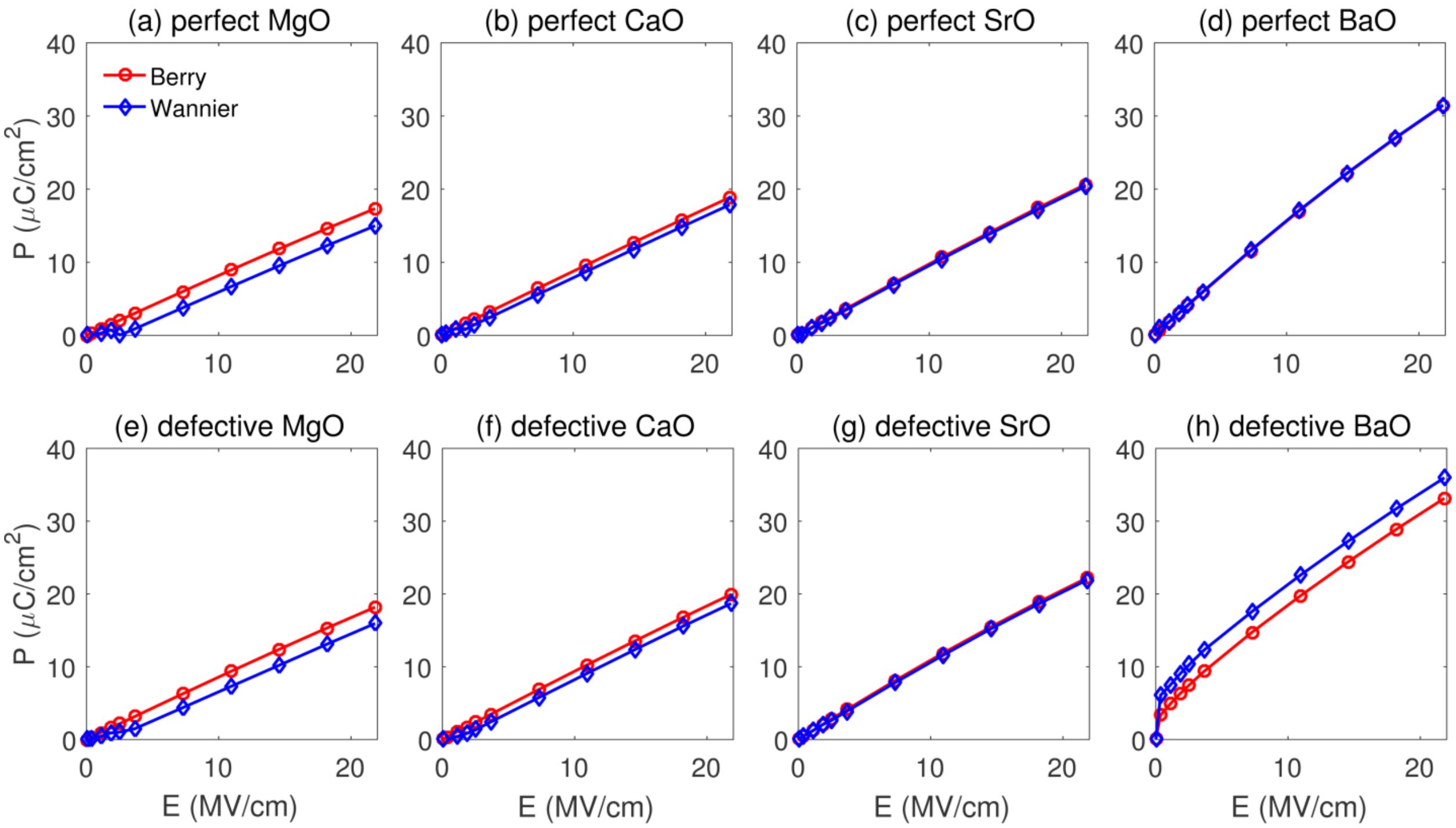

Figure S1. Polarization-electric field relationship computed using Berry phase approach and Wannier centers approach for the perfect crystals (a-d) and defective crystals containing neutral oxygen vacancy (e-h).

At low fields the polarization using Wannier center approach suffers from some oscillations; for example in perfect MgO the polarization slightly decreases although the field increases. We believe that this is due to using a single *k*-point to sample reciprocal space. Low fields result in smaller forces on ions and a denser k-point mesh is needed for accurate computation of forces during the relaxation processes and eventually accurate Wannierization. At high fields Wannier centers polarization becomes a smooth function of the field and its slope becomes very consistent with the slope of the polarization calculated using the Berry phase approach. The magnitudes of the polarizations using the two approaches, however, still have some discrepancies which again originate from the issue of inaccurate low field Wannier centers polarization. Notice that the polarization in Wannier centers approach is calculated in a cumulative fashion, thus an error in the magnitude at low fields will continue to affect the magnitude at high fields. But what really matters in our study is the slope which we use in the paper to compute the site-decomposed polarizability and as Figure S1 shows, the slopes are consistent in both approaches at high fields. In calculating the site-decomposed polarizabilities, we need to assign Wannier centers to each lattice site. Since MgO, CaO, SrO, and BaO are highly ionic materials, the electronic wave functions are highly localized on the lattice sites and we did not face any ambiguity in the assignment. Figure S2 shows the resulting Wannier centers in defective BaO under an electric field of 21.8 MV/cm. Clearly all centers are localized around the lattice sites. In particular, the two centers representing the two electrons trapped in the vacancy are also well-localized on the vacant site.

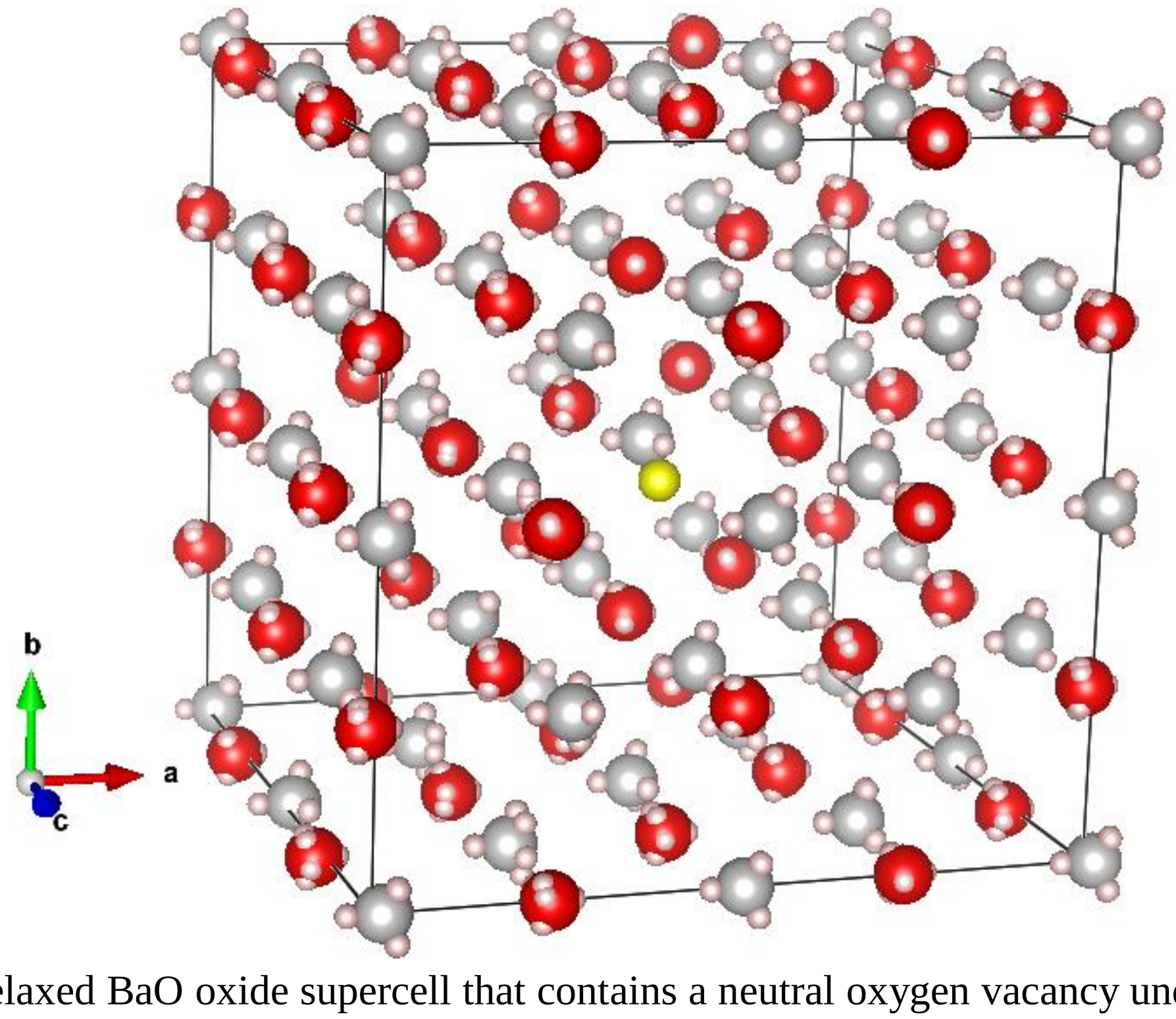

Figure S2. Relaxed BaO oxide supercell that contains a neutral oxygen vacancy under an electric field of 21.8 MV/cm directed along +*x*. Barium and oxygen ions are shown in grey and red, respectively. The centers of maximally localized Wannier functions are shown as small pink spheres. The Wannier centers at the oxygen vacancy site are shown in yellow. This figure was generated using the software VESTA version 3.3.2 [26].

**1.d. Notes on the theoretical approach.**

1) We adopted the notation $G_E$ for the thermodynamic potential resulting from the partial Legendre transform that we introduced in the paper. This is the same notation used for another thermodynamic potential used by Aragonese et al. [27] to study the phase diagram of water under electric field. Here we emphasize the differences between the two potentials. For the purpose of this discussion let's denote the potential introduced here by $G_E$ and the potential of reference [27] by $\tilde{G}_E$. *First,* in the analysis of Aragonese et al. $H_2O$ is considered in integral units without allowing the possibility of dissociation and formation of "defects". Thus, the chemical potentials of the elemental components of water ($\mu_H$ and $\mu_O$) do not appear in the analysis and as such we have, $\tilde{G}_E = N_{H_2O}\mu_{H_2O}$, where $N_{H_2O}$ is the number of water molecules. In our analysis we account for the presence of defects and hence the chemical potentials of oxygen and the elements forming the cation sublattice(s) appear explicitly in our analysis. Thus, $G_E = \sum_k N_k \mu_k$, where the summation is over all cations *k*. *Second*, since defects of water/ice were not considered in reference [27] , the electrochemical work or charge transfer work, $\phi dq$, is not considered in the differential of the internal energy. In our work, although we focus on neutral defects in this paper, we accounted for the electrochemical work for the sake of generality and to furnish the ground for future work that might consider charged defects under electric field.
2) In representing the polarization work we only consider the change in the internal energy of the "material" due to the electric field and hence the term, $\vec{\mathbf{E}} \bullet d(V\vec{\mathbf{P}})$, where *V* is the crystal volume, $\vec{\mathbf{E}}$ is the electric field and $\vec{\mathbf{P}}$ is the macroscopic polarization. But in fact the mere presence of the electric field in the volume *V* results in a change in the energy in that volume regardless whether there is a material or not (see section 3-8 of H. B. Callen text [28]). We chose not to include this "vacuum" energy in our discussion. If we would like to include it, then the polarization work term would be, $\vec{\mathbf{E}} \bullet d(V\vec{\mathbf{D}})$, where $\vec{\mathbf{D}}$ is the electric displacement field. Such term appears in the work of Materlik et al. [29], for example. In a constant volume ensemble both approaches result in the same formation free energy for the defect and in fact the vacuum energy cancels out as it appears in both the perfect and defective crystals free energies. However, if volume changes are considered, then one has to be very careful and choose the expression that correctly describes the situation under investigation.
3) In Berry phase approach to applying electric fields, the DFT code minimizes a functional that already contains the work of polarization However, only the internal energy part of the final result is reliable, whereas the work of polarization part is given on an arbitrary branch of polarization, that is on an arbitrary local minimum of the energy functional as shown in figure 1 of reference [18]. There is no guarantee that the correct polarization branch (correct local minimum) is what is presented in the final results of the DFT code. Even worse, the perfect crystal results could be presented on a polarization branch different than that of the defective crystal results which eliminates any hope of error cancelation. As such we took the field-dependent internal energy directly from the DFT code but we had to post-process the results of the Berry phase calculations in order to identify the correct polarization branch for both the perfect and defective crystal. Only when the correct polarization branches are identified, we can confidently evaluate the correct work of polarization. In fact the same issue arises in calculating the work of polarization using Wannier functions.

4) In references [30,31] a classical sawtooth potential was used to study neutral defects in thin film Si and $TiO_2$, respectively, under external electric field. The results of these two works should be considered under constant electric *displacement* field ($\vec{\mathbf{D}}$) rather than constant electric field ($\vec{\mathbf{E}}$) as in our work. In the sawtooth potential the "macroscopic" electric field should be variable inside the slab system and vary from layer to layer in the material. This was already recognized to some extent in the first work that introduced this approach [32]. Additionally in ref. [31] ,the authors explicitly recognized that the imposed external field is in fact "$\vec{\mathbf{D}}$" and that the electric field is variable inside the slab as clearly marked in Fig. 2 of that work and discussed in the paragraph associated with that the figure. Fixing ($\vec{\mathbf{E}}$) as in our work corresponds to closed-circuit boundary conditions whereas fixing ($\vec{\mathbf{D}}$) corresponds to open circuit boundary condition as discussed in [33]. Under constant ($\vec{\mathbf{E}}$) ensemble a Legendre transform is needed to change the natural variable from dipole to electric field ($\vec{\mathbf{E}}$) as discussed in the manuscript. Under constant $\vec{\mathbf{D}}$ ensemble no Legendre transform is needed and the extensive variable ($(V\vec{\mathbf{D}})$, where $V$ is the volume) is a good natural variable. This explains the appearance of the work of polarization in our formalism and in fact there is no need for such term in constant $\vec{\mathbf{D}}$ ensemble. These ideas are explored in more details in the supplemental materials of [33].

5) We believe that treating defects under external electric fields using modern theory of polarization formulated in constant $\vec{\mathbf{E}}$ as in this work or in constant $\vec{\mathbf{D}}$ [33] is preferable to using the classical sawtooth potential. In the sawtooth potential approach with constant $\vec{\mathbf{D}}$, internal energy includes the polarization effect and it has not been shown how to quantify this polarization effect and compare it to the electrochemical effect. Furthermore, the macroscopic electrostatic potential varies both in the defective crystal and in the perfect crystal and there is no way to align the potential in both cases for bulk (i.e. non-slab) models, making it impossible to extract the electrochemical effect correctly. Envisaging the application of our formalism to assessing charged defects, an ability to resolve and quantify the polarization effects in the defect-free and the defective systems is especially needed. Here, we invoke modern theory of polarization for a point defect in the bulk of a semiconductor for quantifying the polarization effect separately from the electrochemical effect and resolved from the effect of the electric field on internal energy. It is straightforward then to add the electrochemical effects in a post-processing step.

**1.e. Polarizability trends in oxides and gas phase molecules.**

In the manuscript we reported low (zero)-field $\alpha$ for $V_O^{\times}$ to be 20, 46, 139, and 9175 Å$^3$ in MgO, CaO, SrO, and BaO, respectively, increasing with the size of the host lattice. The general trend here that zero-field $\alpha$ of $V_O^{\times}$ increases with increasing the lattice constant of the host oxide is consistent with the same trend found for the $O^{2-}$ ion in these same oxides. The zero-field $\alpha$ for the oxide ion in these four simple oxides ranges between 1.8 and 10.4 Å$^3$ [34]. This trend is also consistent with the increase of zero-field $\alpha$ for gas molecules with the size of the molecule [35]. For example, low (zero)-field $\alpha$ for the highly polar $H_2O$ molecule is 1.47 Å$^3$ ,whereas it is 2.23 Å$^3$for the larger $NH_3$ molecule [35].

**1.f. Kohn Sham states of the neutral oxygen vacancy at zero field.**

At zero field, the center of the doubly occupied Kohn-Sham state of the neutral oxygen vacancy lies below the edge of the DFT conduction band edge by 3.79 eV in MgO, 1.59 eV in CaO, 1.14 eV in SrO, 0.43 eV in BaO, and 0.93 eV in $SrTiO_3$. These values are based in the computational setup described above for the binary oxides and the computational setup described in section 4 below for $SrTiO_3$.

## 2. To smear or not to smear …the dilemma of the neutral oxygen vacancy $V_O^{\times}$ in BaO at zero electric field.

For $V_O^{\times}$ in BaO and under finite electric field we obtain very consistent results whether smearing is applied or not. Surprisingly, a very problematic situation arises for this defect at zero field. Applying smearing (or not) leads to some desirable attributes and some undesirable ones. In particular, applying smearing results in symmetric charge density for the 2 electrons trapped in the vacant site in accordance to the symmetry of the surrounding lattice as shown in Figure S3(a). Such symmetric charge density leads to zero dipole moment for this defect which is anticipated as well. To the contrary, enforcing fixed integer occupations leads to a strangely-looking charge density of these 2 trapped electrons which is not expected based on the symmetry of the surrounding lattice as shown in Figure S3(b). Analyzing Berry phase polarization branches as a function of electric field indicates that this defect has to have a net dipole moment at zero field, which is certainly not justified for such a defect in a simple dielectric such as BaO.

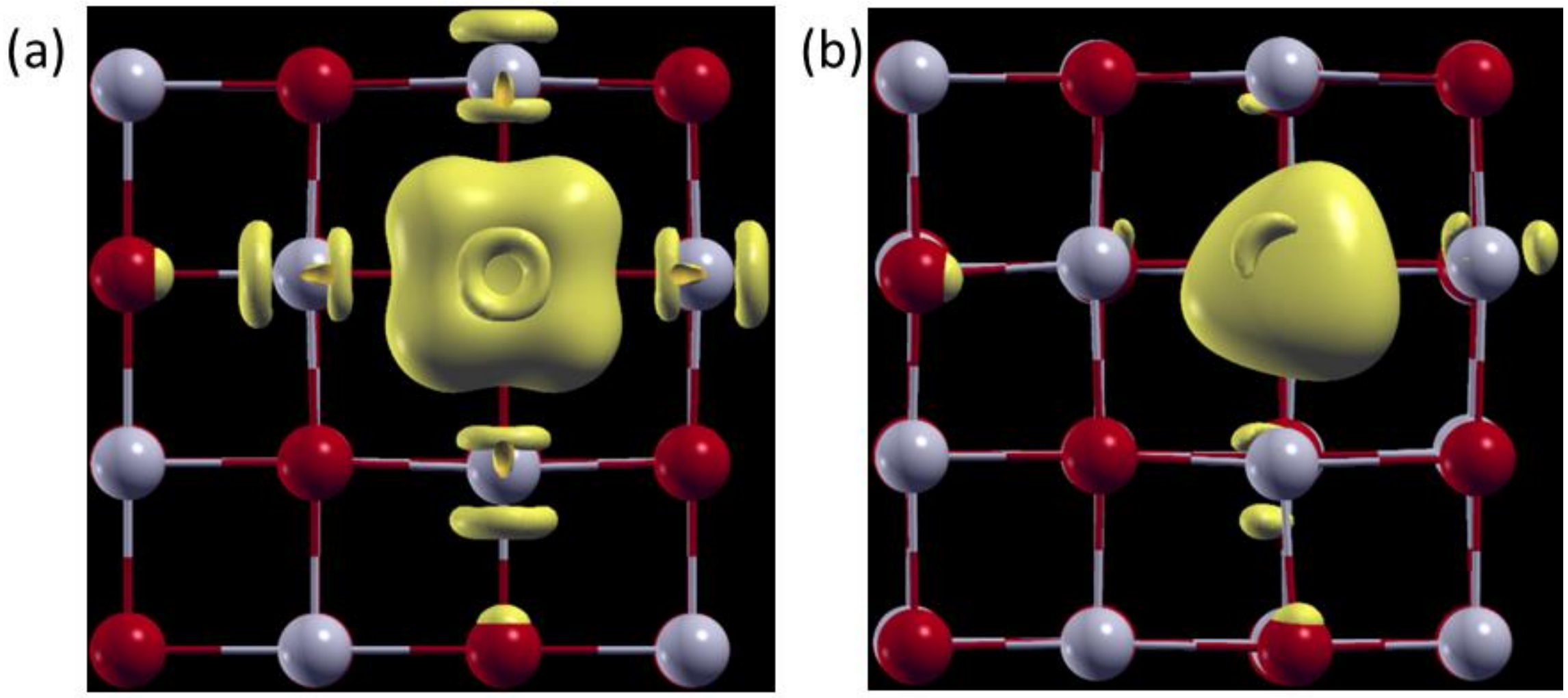

Figure S3. Visualization of the charge density of the two electrons trapped in the neutral oxygen vacancy in BaO as obtained by applying Gaussian smearing (a) and by using fixed occupations and zero smearing (b). Barium and oxygen ions are shown in gray and red, respectively. The yellow isosurfaces represent the charge density of the two electrons trapped in the vacant site and are taken at 15% of the maximum value in each plot. The visualization was generated using the software XCRYSDEN version 1.5.60 [36].

Moreover, the energy of the defective crystal containing $V_O^{\times}$ relaxed using fixed integer occupations is less by 0.16 eV compared to that relaxed using smearing. This energy difference has an important consequence; the inequality $\partial^2 U / \partial(V\vec{\mathbf{D}})^2 \geq 0$ which is dictated by thermodynamic stability considerations [28,37] is only satisfied in the case of fixed occupations. Here $U$ is the internal energy, $V$ is the volume and $\vec{\mathbf{D}}$ is the displacement field. Finally, the DFT solution in the case of fixed occupations is insulator with the defect state localized in the electronic band gap. On the other smearing leads to metallic solution where the defect state mixes with extended conduction band states of BaO. A metallic solution impedes correct application of Berry phase approach to calculate polarization. A summary of this discussion is presented in Table S4 in which green and red are used to highlight desirable and undesirable attributes, respectively. We emphasize again that under finite electric field whether we apply smearing or not, no difficulty arises and the results in both cases are consistent.

Table S4. A comparison between the fixed occupations solution and the smearing solution for the neutral oxygen vacancy in BaO at zero electric field. Desirable attributes are highlighted in green and red is used to highlight undesirable attributes.

| | Fixed integer occupations | Gaussian Smearing |
|---|---|---|
| Total energy and $\partial^2 U / \partial(V\vec{\mathbf{D}})^2 \geq 0$ | 7.00 eV and *adheres to* thermodynamic inequality | 7.16 eV and *does not satisfy* thermodynamic inequality |
| Defect state and DFT solution | localized in-gap state and insulating DFT solution | in part mixes with the conduction band and DFT solution is metallic |
| Symmetry | breaks lattice symmetry | adheres to lattice symmetry |
| Zero field dipole | finite value | zero as expected |

We confirmed that these issues are likely not due to supercell size by repeating the calculations in a supercell made of $3\times3\times3$ conventional cells (108 unit formula). Up to such a large supercell the issues we discussed above persist. Similarly, if we improve the *k*-point sampling in the case of the $2\times2\times2$ supercell, the issues persist. We believe that the origins of these issues are due to the approximation in the exchange correlation functional (PBEsol used here) and the fact that the Kohn-Sham state of $V_O^\times$ at the PBEsol level lie close to the edge of the conduction band. A similar situation arises in analyzing the oxygen atom using DFT. The lower energy solution *does not* possess the experimentally known symmetry of the oxygen atom. Tests using other energy functionals (especially exact exchange functionals) are needed to confirm our hypothesis regarding $V_O^\times$ in BaO. In this study and given that the focus is on electric field effects rather than BaO defect chemistry, we adopted the "*green attributes*" in Table S4 to represent the zero field $V_O^\times$ in BaO.

## 3. Notes on the calculated permittivities.

Our low (zero) field dielectric permittivities for the perfect crystals compare reasonably well with the experimental values in the cases of MgO, CaO, and SrO as shown in Table S5 although these calculations were obtained using a single k-point sampling for the reciprocal space of the supercell as discussed in section 1. In the case of BaO our computed $\varepsilon$ deviates significantly compared to the experimental value. The main reason for this is the sensitivity of the calculated $\varepsilon$ to the lattice constant used in the calculation. Our PBEsol lattice constant of BaO is underestimated compared to the experimental value as in Table S2. A tighter lattice constant, hardens the zone phonon modes and eventually results in underestimation for $\varepsilon$. Such a large sensitivity of $\varepsilon$ to the lattice constant is characteristic of materials at the edge of a ferroelectric transition such as $SrTiO_3$. In fact, BaO can be turned into a ferroelectric material by strain [38]. In spite of this, the qualitative and conceptual discussions in the paper remain valid. Indeed the quantitative discussion also remains valid for a "strained" BaO whose lattice constant is in accordance to our PBEsol lattice constant.

Table S5. Comparison between our DFT-PBEsol calculated values for the static permittivity using applied electric field and Berry phase approach and selected experimental values.

| Oxide | zero field $\varepsilon$ | |
|---|---|---|
| | This work | Experiment |
| MgO | 10.3 | 9.8 [39] |
| CaO | 10.9 | 11.8 [39] |
| SrO | 12.1 | 13.3 [39] |
| BaO | 19.3 | 34 [39,40] |

Finally in the paper we discussed that under electric field the reduction in the dielectric permittivity in perfect and defective oxides is exclusively due to the relaxation (ionic) contribution. In other words, the field hardens the zone center phonon modes but the field effect on the ion-clamped (electronic) permittivity is negligible. To prove this, we show in Figure S4 the components of the dielectric permittivity in all cases considered in this work. Clearly the overall $\varepsilon$ follows the relaxation effect or the ionic contribution. Meanwhile, the ion-clamped (electronic) permittivity remains almost constant as a function of the applied field.

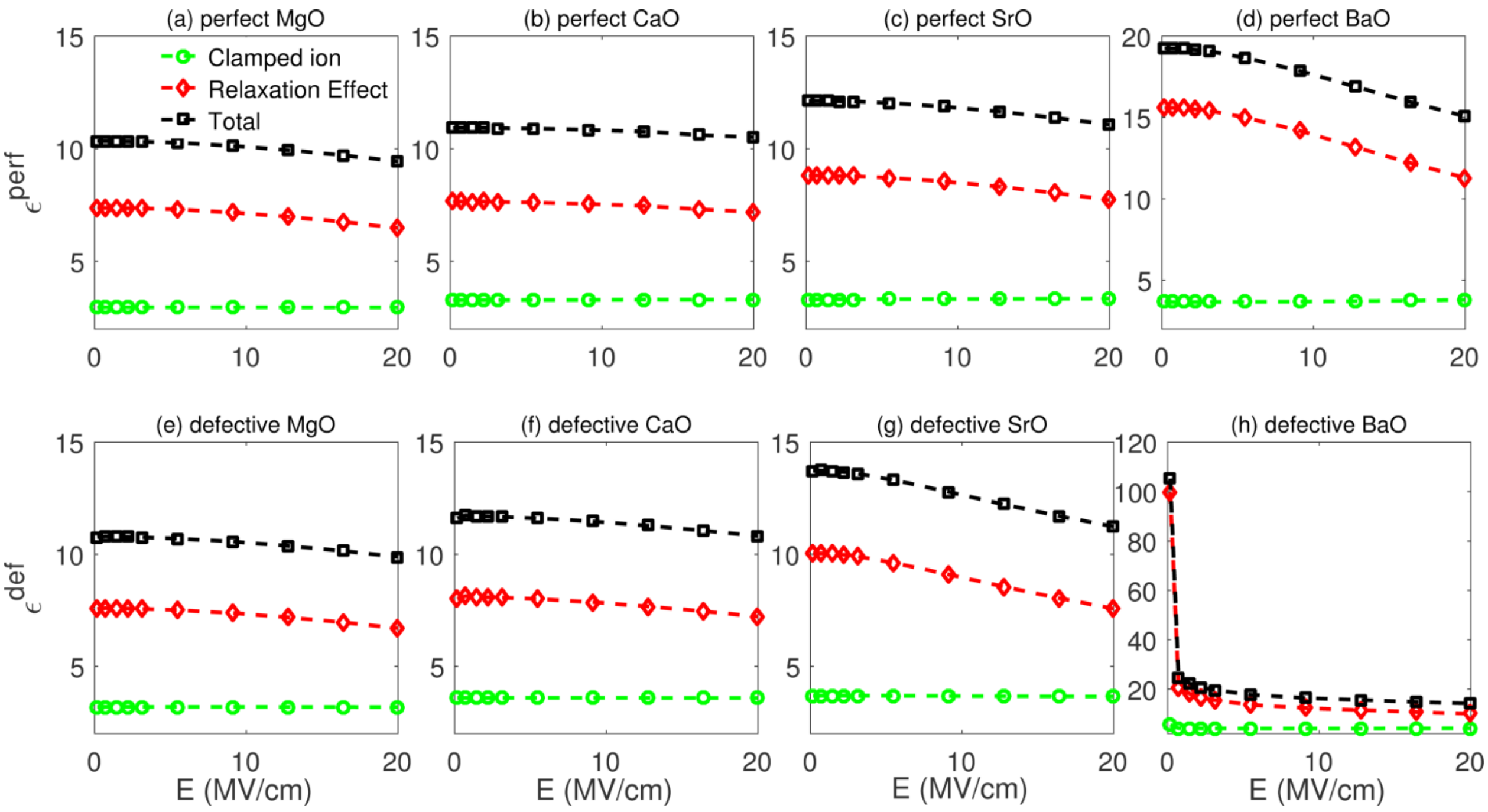

Figure S4. The field dependent static permittivity and its components; clamped ion permittivity and ionic relaxation effect for the perfect crystals (a-d) and defective crystals containing neutral oxygen vacancy (e-h).

### 4. Electric field effect on $V_O^\times$ in $SrTiO_3$.

#### 4.a. Are Born charges in the perfect crystal a good metric for $G_E^{form}$ of $V_O^\times$?

In the manuscript we clarified that Born effective charges $Z^*$ of the cations in the perfect crystal are not necessarily a good metric for the electric field effect $\vec{\mathbf{E}}$ on the free energy of formation $G_E^{form}$ of the oxygen vacancy in this oxide. To demonstrate this point we studied the field effect on the neutral oxygen vacancy $V_O^\times$ in cubic $SrTiO_3$. Based on Born charges of the perfect crystal cations shown in Table S6, one would imagine that $\Delta G_E^{form}$ in the case of $SrTiO_3$ would be lower than $\Delta G_E^{form}$ in BaO under the action of the field. However, this is not the case as shown in Figure S5 (a).

Table S6. Born effective charges of the cations surrounding $V_O^\times$ in five oxides. The charges were calculated using density functional perturbation theory [41] except in the case of $SrTiO_3$ with the functional PBEsol+$U_{Ti}$ which was calculated using the application of electric field.

| Cation in oxide | Functional | Born Charge | Born Charge / formal Charge |
|---|---|---|---|
| $Mg^{2+}$ in MgO | PBEsol | +2.0 | 1.00 |
| $Ca^{2+}$ in CaO | PBEsol | +2.3 | 1.15 |
| $Sr^{2+}$ in SrO | PBEsol | +2.4 | 1.20 |
| $Ba^{2+}$ in BaO | PBEsol | +2.7 | 1.35 |
| $Ti^{4+}$ in $SrTiO_3$ | PBEsol | +7.3 | 1.83 |
| $Ti^{4+}$ in $SrTiO_3$ | PBEsol+$U_{Ti}$ | +6.4 | 1.60 |

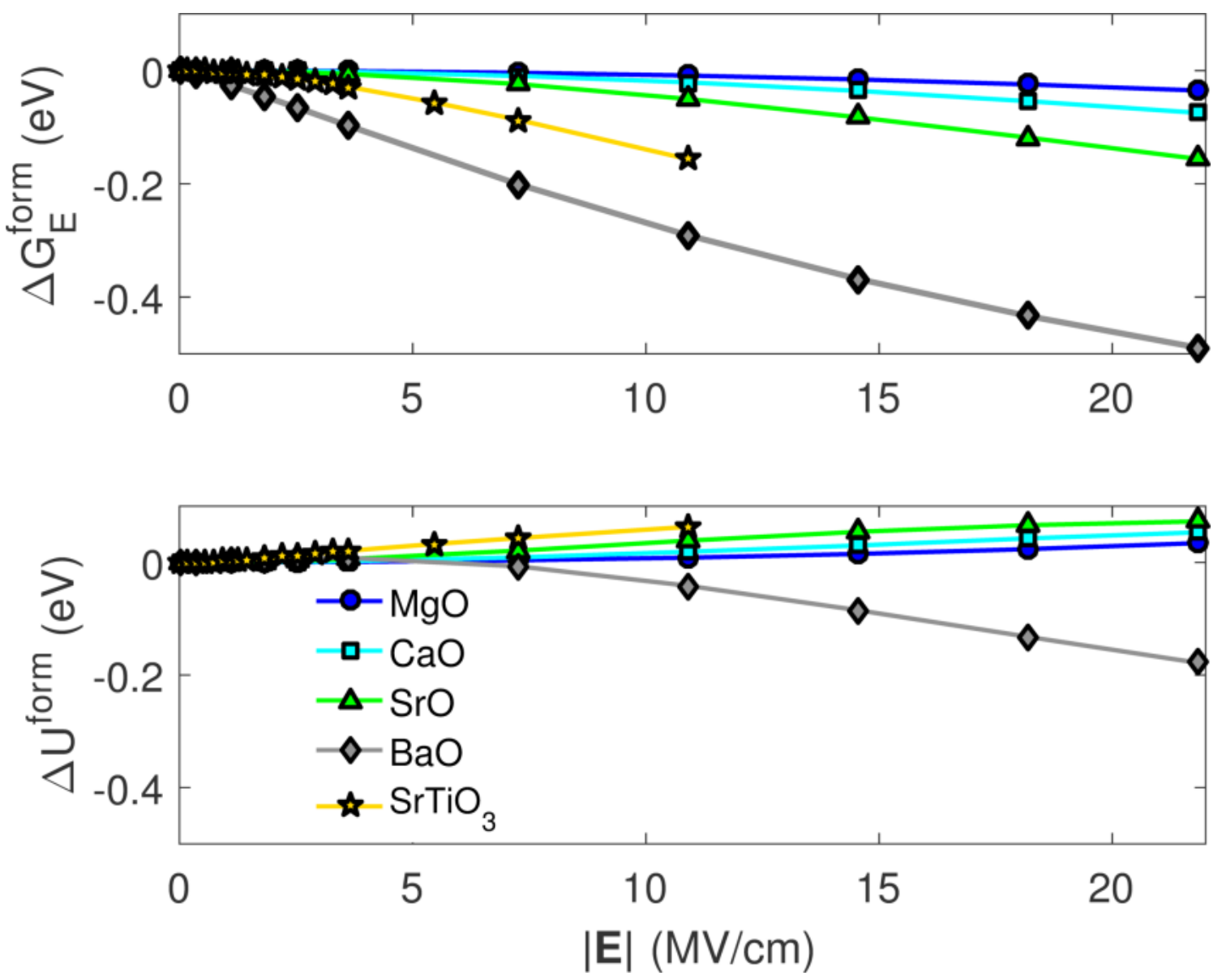

Figure S5. (a) Relative electric Gibbs free energy of formation and (b) relative formation energy of the neutral oxygen vacancy as a function of electric field.

It is true that the unusually large Born charge of $Ti^{4+}$ cation in $SrTiO_3$ implies that the lattice is highly polarizable and even more polarizable than BaO. In fact, our explicit calculations of the polarization as a function of electric field of the perfect crystals shown in Figure S6(a) and defective crystals shown in Figure S6(b) are consistent with the conclusions based on Born charges. However, what governs the work of polarization term is the net difference between the perfect and defective crystals. This net difference is represented in Figure S6(c) which shows the dipole moment of $V_O^\times$ as a function of electric field. The net dipole moment of $V_O^\times$ in BaO oxide is larger than in $SrTiO_3$ and, consequently, the work of polarization is able to lower $\Delta G_E^{form}$ in BaO more than in $SrTiO_3$. In summary, Born effective charges of the cations in the perfect crystal are not necessarily predictive of the behavior of $\Delta G_E^{form}$ of $V_O^\times$ under electric field.

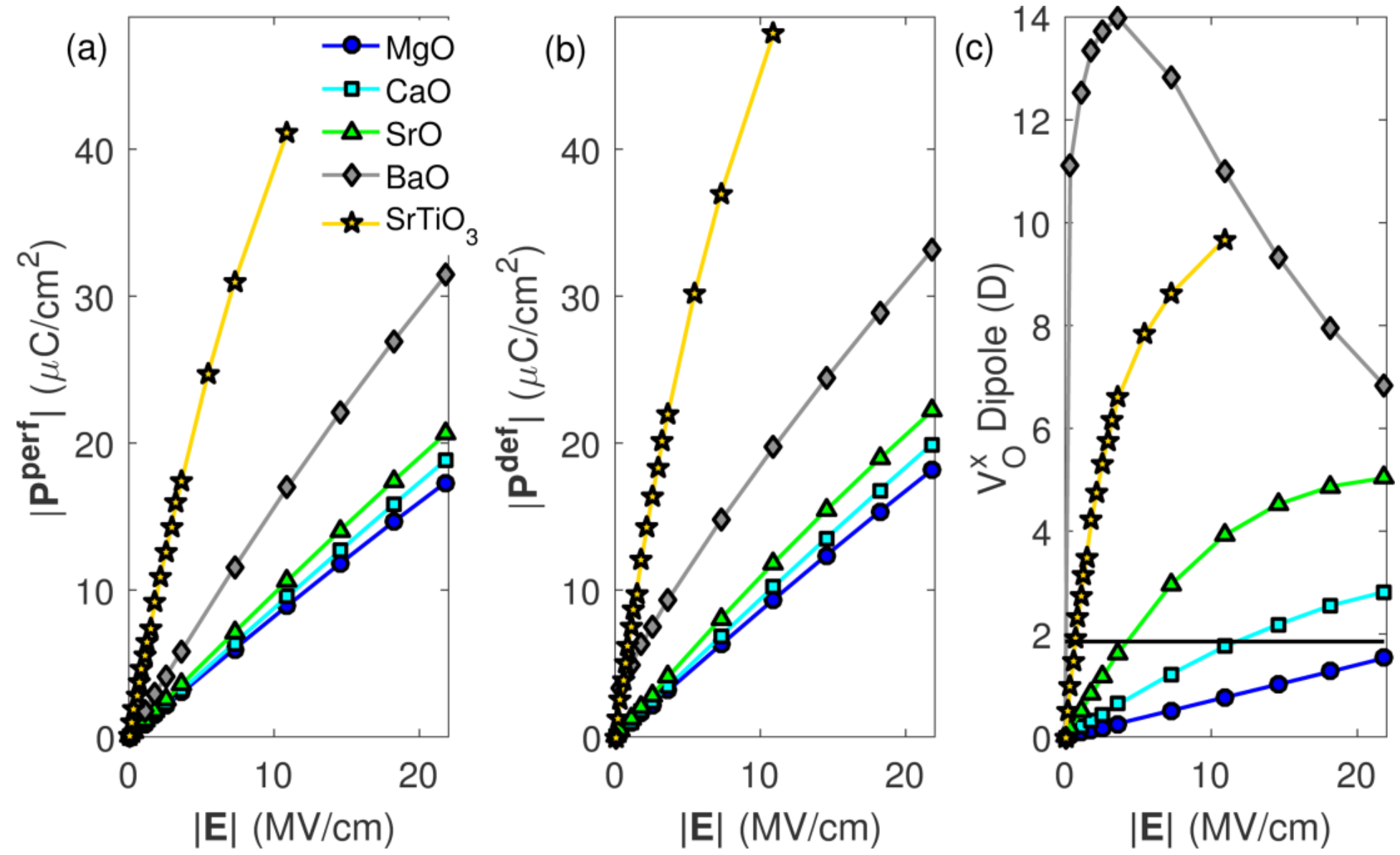

Figure S6. Polarization of (a) the perfect crystals and (b) defective crystals as a function of electric field. (c) The field dependence of the dipole moment of the neutral oxygen vacancy. For comparison the zero-field dipole moment of the gas-phase water molecule, 1.86 D [42], is indicated by the black horizontal line.

**4.b. Computational details for $SrTiO_3$ under electric field.**

In this work we adopted the cubic centrosymmetric structure of $SrTiO_3$ to serve the comparison with the centrosymmetric binary oxides studied in this work. Most of the computational details for $SrTiO_3$ are similar to those of the simple binary oxides detailed in section 1.a. and 1.b. above. Here we provide the exceptions. Ti pseudopotential was selected from GBVR library [4], `ti_pbesol_v1.4.uspp.F.UPF`. The exchange correlation was represented using PBEsol [3] functional equipped with an on-site Coulomb interaction term ($U$) of 5 eV on Ti 3d states. The $U$ was applied using the method of Cococcioni and De Gironcoli [43]. Using this setup we obtained a lattice constant of 3.922 Å, bulk modulus $K$ of 174 GPa, and electronic band gap of 2.7 eV. The experimental values of these properties are 3.900 Å for lattice constant [44], 179 GPa for bulk modulus [45], and 3.25 eV for band gap [46].

Electric field calculations of the perfect and defective crystals were performed in supercells made of 2×2×2 conventional cells (8 unit formula). Reciprocal space was sampled using 2×2×2 shifted Monkhorst-Pack [9] *k*-point grid. The electric field was applied along the Ti-$V_O^\times$-Ti direction which is also coincident with [001] direction. We attempted applying it perpendicular to this direction but we observed a structural phase transition taking place in that case similar to the one observed in reference [47]. So in order not complicate the analysis we focused here on applying the field parallel to Ti-$V_O^\times$-Ti direction. $SrTiO_3$ admits two types of neutral oxygen vacancies. A polaronic one in which the two electrons reminiscent of the oxide ion are localized on the neighboring Ti cations, and a color-center-like type in which the two electrons are trapped in the vacant site [48]. To facilitate the comparison with the alkaline-earth-metal binary oxides, we restricted the analysis here to the color-center-like oxygen vacancy.

It is worth mentioning that DFT+*U* is necessary to achieve the localization of the two electrons in the vacant site and a corresponding insulating solution. Pure DFT (that is pure PBEsol) would result in a metallic solution in which the two electrons mix with the extended conduction band states.

## 5. Why does the behavior of $V_O^\times$ in BaO look different?

The results of BaO look different compared to the rest of the binary oxides considered here. But this difference is physically sound and points to the interesting interplay between defects, phonons, and external electric fields. The family of binary oxides we studied was shown to exhibit ferroelectric behavior upon applying tensile strain [38]. This is accomplished by softening some phonon modes via strain. BaO was the easiest among this family to turn into ferroelectric thanks to its larger lattice constant. In a similar fashion, the oxygen vacancy also softens some phonon modes as evidenced from the larger dielectric permittivity of the defective crystal in all cases (Fig. 4 of the paper). Since BaO has the largest lattice constant among the series, introducing the vacancy in this material brings it to the verge of being ferroelectric as evidenced from the very large dielectric permittivity at low field of the defective BaO (Fig. 4 (b) of the paper). This, in turn, enhances the polarizability of defective BaO at low-fields significantly compared to the rest of the binary oxides, and leads to the very large increase in the vacancy dipole moment between 0 and 3 MV/cm as shown in Fig. 2( a) of the paper. In fact, between 0 and 3MV/cm BaO is no different in behavior compared to the other binary oxides. Simply, its vacancy is much more polarizable since the host in which the vacancy is embedded has the larger lattice constant and as such is closer to being ferroelectric. We believe that the same effect can be induced in the other oxides as well via tensile hydrostatic strain. For example, SrO at the lattice constant of BaO can exhibit similar very large vacancy dipole moment at low fields. Now why at around 3 MV/cm the dipole moment of the vacancy in BaO starts to decrease? The reason is that the phonon modes that start softer at zero field will harden at a faster rate as the field increases [49]. Thus, the phonon modes of defective BaO will harden under the field at a faster rate than the phonons of perfect BaO. As such at a certain value for the field (3MV/cm in BaO) the perfect crystal will have effectively softer modes compared to the defective crystal. This leads to decrease in the dipole moment of $V_O^\times$ (Fig. 2a in the paper), negative polarizability of the defect (Fig. 2b in the paper), larger dielectric constant for the perfect crystal (Fig. 4 in the paper), and decreasing relative formation energy ($\Delta U^{form}$) for the defect (Fig. 1b in the paper). Since $\Delta U^{form}$ became decreasing in BaO, $\Delta G_E^{form}$ become even more decreasing as in Fig. 1a of the paper.

## Supplemental references